\documentstyle[12pt,axodraw]{article}
\newif\ifall\alltrue
\newif\ifprivate\privatetrue
\privatefalse % privates nicht compilieren

\def\ignore#1{}
\def\deltabar{\delta\!\!\!^{-}}
\def\hbar{h\!\!\!\!^{_-}\,}

\begin{document}
%%%%%%%%%%%%%%%%%%%%%%%%%%%%%%%
%         Title-Page          %
%%%%%%%%%%%%%%%%%%%%%%%%%%%%%%%

\ifall

\hspace*{11cm}LMU 93-18

\hspace*{11cm}November 1993

\begin{center}       \vspace*{3cm}
{\LARGE Gluon Mass from Instantons}               \\[3cm]
  {\bf Marcus Hutter\footnotemark}                \\[2cm]
  {\it Sektion Physik der Universit\"at M\"unchen}\\
  {\it Theoretische Physik}                       \\
  {\it Theresienstr. 37 $\quad$ D-80333 M\"unchen}\\[2cm]
\end{center}
\footnotetext{E--Mail: hutter@hep.physik.uni--muenchen.de}

\begin{abstract}
The gluon propagator is calculated in the instanton background
in a form appropriate for extracting the
momentum dependent gluon mass.
In background-$\xi$-gauge we get for the mass $400$MeV
for small $p^2$ independent of the gauge parameter $\xi$.
\end{abstract}
\newpage
\fi

%%%%%%%%%%%%%%%%%%%%%%%%%%%%%%%
\paragraph{1. Introduction} \hfill
%%%%%%%%%%%%%%%%%%%%%%%%%%%%%%%

\ifall
In the following we will calculate the gluon propagator for
different gauges. In background $\xi=1$ gauge it has the simple form
$ S_{\mu\nu}^{ab}=\delta^{ab} g_{\mu\nu} / (p^2-M(p)^2) $.
In section 2 we will extract the inverse propagator from
the terms quadratic in the fluctuations around a background
field. In section 3 we will average this expression over
relevant background fields and expand it for large momentum in a
way to get the gluon mass $M(p)$. In section 4 we explicitly
calculate $M_1=M(p=\infty)=420$MeV in the instanton background.
We will also see that there are further terms which do
not vanish at large momentum but are more difficult to
calculate. To get reliable results for large as well as for
small momentum we make in section 5 an cluster expansion in
the instanton density. For this purpose we need the gluon propagator
in the 1 instanton background. The relevant formulas to construct
this propagator are listed in section 6. Because it is a bit lengthy
we will restrict ourself in section 7 to the case of small momentum.
Till now as far as I know the instanton background is only treated
at the classical level except for a calculation of the one loop
action in the instanton background.
This work should be seen as a first step to calculate
gluonic quantum fluctuations around the instanton background
and the simplest object to play with is the propagator.
For high energies there are tools like operator product expansion
with instantons included indirectly in the condensates \cite{SVZ}.
This is another reason for the restriction to small momentum.
\fi

%%%%%%%%%%%%%%%%%%%%%%%%%%%%%%%
\paragraph{2. Gluon Propagator} \hfill
%%%%%%%%%%%%%%%%%%%%%%%%%%%%%%%

\ifall
The first task to obtain a formal expression for the
gluon propagator in a background field
is to expand ${\cal L}_{QCD}[\bar{A}+B]$ in the fluctuations
$B_\mu^a$ around our background $\bar{A}_\mu^a$.
The term quadratic in $B_\mu^a$ is then by definition
the inverse gluon propagator.
For $\bar{A}_\mu^a$ we will later use our instanton gas.
We will use the QCD-Lagrangian
$$
  {\cal L}_{QCD}= {1\over 4g^2} G_{\mu\nu}^a G^{\mu\nu}_a \quad,\quad
  G_{\mu\nu}^a(A)=\partial_\mu A_\nu^a -
             \partial_\nu A_\mu^a +
             f_{abc} A_\mu^b A_\nu^c
$$
where we have rescaled the fields in such a way that the
coupling-constant-dependence is in front of $\cal L$.
We will entirely work in Euclidian space with metric
$\delta_{\mu\nu}$ instead of $g_{\mu\nu}$ because instantons do
not make sense in Minkowski space. At the very end we can
simply rotate back to Minkowski space.
With these conventions
$$
  g^2{\cal L}_{QCD}(\bar{A}+B) =
    {1\over 4} G_{\mu\nu}^a(\bar{A}+B)
    G^{\mu\nu}_a(\bar{A}+B) =
$$ $$
  = {1\over 4} \overbrace{ \bar{G}_{\mu\nu}^a
                  \bar{G}^{\mu\nu}_a }^{O(B^0)} +
  \overbrace{ B_\mu^a\bar{D}_\nu^{ab}
              \bar{G}_a^{\mu\nu} }^{O(B^1)} +
  {1\over 2} \overbrace{ B_\mu^a (
    -\bar{D}_\rho^{ac}\bar{D}_\rho^{cb}\delta_{\mu\nu}
    -2f_{acb}\bar{G}_{\mu\nu}^c+
    \bar{D}_\mu^{ac}\bar{D}_\nu^{cb} ) B_\nu^b }^{O(B^2)} +
$$
\begin{equation}\label{LQCD}
  + \underbrace{ f_{abc}B_\mu^b B_\nu^c\bar{D}_\mu^{ad}
                           B_\nu^d }_{O(B^3)} +
  {1\over 4} \underbrace{ f_{abc}B_\mu^b B_\nu^c f_{ade}B_\mu^d
                          B_\nu^e }_{O(B^4)} +
  \partial_\mu(\ldots) \quad,
\end{equation}
where
$ \bar{D}_\mu^{ab}=\partial_\mu\delta_{ab}+
    f_{acb}\bar{A}_\mu^c
$
is the covariant derivative with $\bar{A}$ inserted instead of
$A$; similarly $\bar{G}_{\mu\nu}^a=G_{\mu\nu}^a(\bar{A})$.
As usual the terms which are total derivatives disappear
after integrating $\cal L$.
We see that if $\bar{A}$ solves the
equation of motion the linear term vanishes, as it should be.
If we choose background gauge $\bar{D}_\nu^{ac}B_\nu^c=0$ the last
term of the $O(B^2)$-contribution vanishes.
Now we can read the inverse gluon propagator from the terms
quadratic in $B$ (from now on we will omit the bars over $A$, $G$ and
$D$ because the unbared objects won't be needed further):
\begin{equation}
  (S^{-1})_{\mu\nu}^{ab}={1\over g^2}
  (-D_\rho^{ac}D_\rho^{cb}\delta_{\mu\nu}
  -2f_{acb}G_{\mu\nu}^c) \quad.
\end{equation}
We will also omit the $1/g^2$ in front of the propagator
which is a result of the rescaling of fields anyway.
For further manipulations some abbreviations are useful:
$$
  G_{\mu\nu}=F^c G_{\mu\nu}^c \quad,\quad
  A_\mu=F^c A_\mu^c \quad,
$$ $$
  (F^c)_{ab}=i f_{acb} \quad,\quad
  [F^a,F^b]=i f_{abc}F^c \quad,\quad
  tr_c F^a\!F^b=N_c \delta^{ab} \quad,\quad
  N_c=3 \quad,
$$ $$
  (\hat{P}_\mu)^{ab}=iD_\mu^{ab} \quad,\quad
  \hat{p}_\mu=i\partial_\mu \quad,\quad
  \hat{P}_\mu=\hat{p}_\mu+A_\mu \quad,
$$
\begin{equation}
  \hat{p}_\mu X=[\hat{p}_\mu,X]+X \hat{p}_\mu = i(\partial_\mu X)+X
\hat{p}_\mu \quad.
\end{equation}
The last equation has only been quoted to show that
$\hat{p}$ and $\hat{P}$ will be used in operator sense.
$F^a$ are the generators in adjoint representation and
$f_{abc}$ are the structure constants of the color gauge group $SU(N_c)$.
With these abbreviations we can now write
$$
  S^{-1}_{\mu\nu} = \hat{P}^2\delta_{\mu\nu}+2iG_{\mu\nu}
  = (\hat{p}^2+\hat{p}\!\cdot\!A+A\!\cdot\!\hat{p}+A^2)
    \delta_{\mu\nu}+2iG_{\mu\nu}
  = (S_0^{-1}+V)_{\mu\nu} \quad,
$$
\begin{equation}\label{Sinv}
  S^0_{\mu\nu}=\delta_{\mu\nu}/{\hat{p}^2} \quad,\quad
  V_{\mu\nu}=(A^2+\hat{p}\!\cdot\!A+A\!\cdot\!\hat{p})
  \delta_{\mu\nu}+2iG_{\mu\nu} \quad.
\end{equation}
$S_0$ is the free gluon propator with no background and $V$
can be interpreted as an interaction potential caused
by the background.
The QCD-Lagrangian in the background \ref{LQCD} can be written
in another form more suitable for ordinary perturbation theory:
\begin{equation}
  g^2{\cal L}_{QCD}={1\over 4}G_{\mu\nu}^a(B)G^{\mu\nu}_a(B)
  +{1\over 2}B_\mu^a V_{\mu\nu}^{ab}B_\nu^b
  +f_{abc}f_{aed}B_\mu^b B_\nu^c B_v^d\bar{A}_e^\mu
\end{equation}
The terms independent and linear in $B$ has been omitted
because for a given background the first term is irrelevant and
the second can be deleted by shifting $B$. To determine the
weight of a specific background they are of course the most
important terms. The Feynman graphs of the additional terms
(with $g$ recovered) are depicted in  figure \ref{Fey}.
It should be noted that for small coupling constant $g$ the
second graph can be treated perturbatively but not the first.
So our main concentration lies on the first term.

\begin{figure}\label{Fey} \fboxsep=8mm
\framebox[\textwidth]
{
\unitlength=1mm
\SetScale{2.835}
\SetWidth{0.2}
\begin{picture}(68,42)
\put(28,14){\circle{10}}
\put(28,14){\makebox(0,0)[cc]{A}}
\put(68,14){\makebox(0,0)[lc]{$=g f_{abc}f_{aed}A^e_\mu\delta_{\mu\rho}$}}
\put(4,14){\makebox(0,0)[rc]{$\mu,b$}}
\put(43,26){\makebox(0,0)[lc]{$\nu,c$}}
\put(43,2){\makebox(0,0)[lc]{$\rho,d$}}
\put(28,37){\circle{10}}
\put(28,37){\makebox(0,0)[cc]{V}}
\put(4,37){\makebox(0,0)[rc]{$\mu,a$}}
\put(52,37){\makebox(0,0)[lc]{$\nu,b$}}
\put(68,37){\makebox(0,0)[lc]{$=V_{\mu\nu}^{ab}$}}
\Gluon(23,37)(5,37){2}{4}
\Gluon(33,37)(51,37){2}{4}
\Gluon(23,14)(5,14){2}{4}
\Gluon(32,17)(41,26){2}{3}
\Gluon(32,11)(41,2){2}{3}
\end{picture}
}
\vspace{-2ex}
\caption{\it Feynman rules for gluons in a background}
\end{figure}
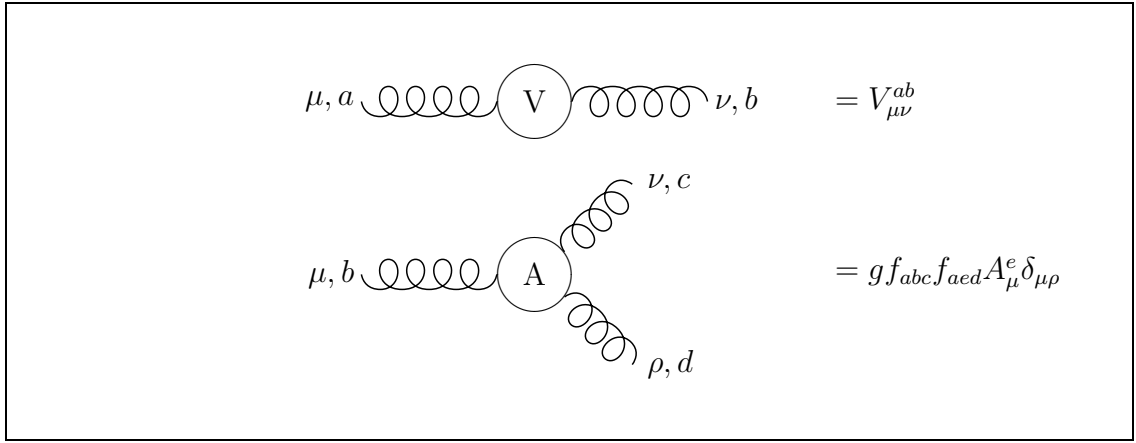
\fi

%%%%%%%%%%%%%%%%%%%%%%%%%%%%%%%
\paragraph{3. Propagator in Statistical Background} \hfill
%%%%%%%%%%%%%%%%%%%%%%%%%%%%%%%

\ifall
The next step is to use some approximation scheme to calculate
the propagator
\begin{equation}
  S=(S_0^{-1}+V)^{-1}=S_0(1\!\!1+T)^{-1} \quad,\quad T=VS_0 \quad .
\end{equation}
For large momentum $p$, $S_0$ and therefore $T$ are small and we
can expand $S$ in powers of $T$:
\begin{equation}
  S=S_0(1\!\!1-T+T^2-T^3+\ldots) \quad .
\end{equation}
Note that $S(x,y)=\langle x|S|y\rangle$ is generally not
invariant under translations and rotations
because the background $A_\mu$ and therefore $T$ are not.
Actually we are not interested in the propagator for a particular
background configuration, but only in the average over all relevant
configurations. Here we do not mean the functional integration
over quantum fluctuations around the empty vacuum, but fields
other than the perturbative $A_\mu\!=\!0\,$-vacuum which
minimize the action $\int{\cal L}\,dx$.
\begin{equation}\label{Sav}
  \overline{S}=S_0(1\!\!1-\overline{T}+\overline{T^2}-\overline{T^3}
                    +\ldots) \quad,
\end{equation}
where the bar denotes averaging over relevant configurations.
Details are specified below. If the background
is statistically invariant under translations then
$\overline{S}$ is translationally invariant and therefore
diagonal in momentum space
$  \overline{S}(p,q)=\langle p|\overline{S}|q\rangle =
   \overline{S}(p)\deltabar(p-q)$.
If we would evaluate the terms in the series
we would get an expansion of $\overline{S}$ in the form
\begin{equation}\label{Sa2}
  \overline{S}(p)=1/p^2+c_1/p^4+c_2/p^6+\ldots
\end{equation}
and the pole remains at zero --- and gets even worse with higher terms.
What we want is $\overline{S}$ in a form like
$\overline{S}(p)=(p^2+M(p)^2)^{-1}$ with $M(p)$ bounded for large
and small $p$ and interpreted as momentum dependend
gluon mass\footnote{
The $"+"$ in front of $M$ will change to the more familiar $"-"$ when
we rotate back from Euclidian space to Minkowski space.}.
So let's invert (\ref{Sav})
\begin{equation}
  \overline{S}^{-1}=\hat{p}^2+M(\hat{p})^2=(1\!\!1-\overline{T}+
  \overline{T^2}-\overline{T^3}+\ldots)^{-1}S_0^{-1}
\end{equation}
and expand it once again in $T$. Without averaging this would
just be a geometrical series which, expanded, would give the original
formula $S^{-1}=S_0^{-1}+V$. With averaging the different
terms are now in
no relation and expanding and sorting with respect to powers of
$T$ yields
\begin{eqnarray}\label{Sai}
  \overline{S}^{-1} &=& ( 1\!\!1+\overline{T}
   - (\overline{T^2}-\overline{T}^2)
   + (\overline{T^3}-\overline{T}\,\overline{T^2}
     -\overline{T^2}\,\overline{T}+\overline{T}^3) )S_0^{-1}
   + O(T^4)
\\
   &=& S_0^{-1}+\overline{V}
   -(\overline{VS_0V}-\overline{V}S_0\overline{V}) + \ldots
\nonumber \\
   &=& S_0^{-1}+M^2 ,\quad M^2=M_1^2-M_2^2+\ldots ,\quad
   M_1^2=\overline{V} ,\quad
   M_2^2=\overline{VS_0V}-\overline{V}S_0\overline{V} .\nonumber
\end{eqnarray}
In the next section we introduce the concepts of
instanton gas calculation to determine $M_1$, which is
actually very simple.

\fi
%%%%%%%%%%%%%%%%%%%%%%%%%%%%%%%
\paragraph{4. First Order Gluon Mass} \hfill
%%%%%%%%%%%%%%%%%%%%%%%%%%%%%%%

\ifall
The classical approximation to a quantum theory in the language
of path integrals is to consider only the configurations which
minimize the classical Euclidian action $S$. In QED there is only
one minimum, namely $A_\mu=0$, but in QCD there are other local
minima, called instantons:
\begin{equation}
  A_{I\mu}^a (x)=O_I^{ab}\eta^I_{b\mu\nu} {(x-z_I)_\nu\over (x-z_I)^2}
                      {2\rho^2\over (x-z_I)^2+\rho^2} \quad,
\end{equation}
$$
  \eta^I_{a\mu\nu}=\overline{\eta}_{a\mu\nu}=
  \epsilon_{a\mu\nu 4}-{1\over 2}\epsilon_{abc}\epsilon_{bc\mu\nu}
  \quad\mbox{for instantons} \quad,
$$ $$
  \eta^I_{a\mu\nu}=\eta_{a\mu\nu}=
  \epsilon_{a\mu\nu 4}+{1\over 2}\epsilon_{abc}\epsilon_{bc\mu\nu}
  \quad\mbox{for anti-instantons} \quad.
$$
$A_{I\mu}^a$ is an (anti)instanton of size $\rho$ at position $z_I$
in singular gauge and orientation $O_I$, where $O_I$ is a
rotation matrix in the adjoint representation of $SU(N_c)$.
$\eta_{a\mu\nu}$ are the 't Hooft-symbols with some properties
given in the appendix of \cite{tHo}. Whereas the single instanton is an
exact solution of the equation of motion $D_\mu^{ab}G_{\mu\nu}^b=0$,
the multi-instanton-field $A=\sum_I A_I$ minimizes $S$
only approximatly, but nevertheless gives a significant contribution
to the functional integral.
Quantum fluctuations around the instanton renormalize the
action $S=8\pi^2/g^2$ and make small instantons less important
because the QCD coupling constant $g$ decreases for small distances
\cite{tHo}.
Repulsion between instantons do not allow too large instantons
\cite{Dya}, \cite[A]{Shu}.
So there is a narrow region of allowed values for the
instanton radius and I will
use an average radius $\rho$ in my calculations. The second
parameter we need is the instanton density $n$
which must be determined experimentally from the
gluon  condensate \cite{SVZ}.
The ratio $L_0/\rho$ is calculated in \cite{Dya}, \cite{Shu}:
\begin{equation}
   n= N/V_4 = 1/L_0^4 =
   {1\over 32\pi^2}\langle G_{\mu\nu}^a G^{\mu\nu}_a\rangle =
   (200\mbox{MeV})^4_{exp.} \quad,\quad
   L_0/ \rho=3.2_{theor.}
\end{equation}
Now we plug the N-(anti)instanton-field
\begin{equation}
  A=\sum_{I=1}^N A_I
\end{equation}
in our expressions for the gluon propagator (\ref{Sai}).
To do this we must
average some functions $f(A)$ of $A$ like in $\overline{V}$.
We will use the instanton gas approximation without interactions
between instantons. The whole effects of interactions are summarized
in the values of $L_0$ and $\rho$. So the position and orientation
are equally distributed and independent for different instantons:
\begin{equation}
  \overline{f(A)}=\prod_{I=1}^N {1\over V_4}
  \int d^4\!z_I\int dO_I\,f(A) \quad,
\end{equation}
where ${1\over V_4}\int dz_I$ averages over the position of
instanton $I$ contained in a large box of volume $V_4$.
$\int dO_I$ is the Haar-measure and averages over the group of
orientations
$$
  \int dO\,1=1 \quad,\quad
  \overline{O^{ab}}=\int dO\,O^{ab}=0 \quad,\quad
$$
\begin{equation}
  \overline{O^{ab}O^{cd}}=\int dO\,O^{ab}O^{cd} =
    {1\over N_c^2-1}\delta^{ac}\delta^{bd} .
\end{equation}
We now can easily calculate
$  \overline{V}_{\mu\nu} =
    (\overline{A^2}+\hat{p}\cdot \overline{A}+\overline{A}\cdot \hat{p})
    \delta_{\mu\nu}+2i\overline{G}_{\mu\nu} \;:
$
\begin{equation}\label{Aav}
  \overline{A_\mu^a}=N\overline{A_{I\mu}^a}
   =N\overline{O^{ab}}\cdot\ldots = 0 \quad\mbox{and}\quad
   \overline{G_{\mu\nu}^a}=0
\end{equation}
because $G_{\mu\nu}^a$ is antisymmetric in the Lorentz indices
and Lorentz invariant\footnote{Precisly we should say $O(4)$ invariant
because we are in Euclidian space.}
after averaging, which is impossible.
$$
   \overline{A^2}=\sum_{I=1}^N\overline{A_I^2}
     + \sum_{I\neq J}\overline{A_I A_J} =
   N\overline{A_I^2}+N(N-1)\overline{A}_I\overline{A}_J =
   N\overline{A_I^2} \quad,
$$
where we have used the fact that different instantons are independent
and (\ref{Aav}).
Inserting now the specific form of the instanton we yield
\begin{eqnarray}
   (\overline{A^2})_{ab}
&=&
   {N\over V_4}\int dz_I\int dO\,
   if_{acd}A_{I\mu}^c\,if_{deb}A_{I\mu}^e \nonumber\\
&=&
   -{N\over V_4}f_{acd}f_{deb}\overline{O^{cf}O^{eg}}
   \eta^I_{f\mu\nu}\eta^I_{g\mu\rho}\int d^4\!z{z_\nu z_\rho\over z^4}
   \left( {2\rho^2\over z^2+\rho^2} \right)^2 \nonumber\\
&=&
   {12\pi^2 N_c\over N_c^2-1}n\rho^2\delta_{ab} \quad,\quad
   n={N\over V_4}=\mbox{instanton density}
\end{eqnarray}
where we have used $f_{acd}f_{dcb}=-N_c\delta_{ab}$ and
$\eta^I_{a\mu\nu}\eta^I_{a\mu\rho}=3\delta_{\nu\rho}$.
Inserting this into $M_1^2=\overline{V}=\overline{A^2}$
we get
\begin{equation}\label{M1}
   M_1=\sqrt{12\pi^2 N_c\over N_c^2-1}(\rho/L_0)L_0^{-1}
   \approx 2.1 L_0^{-1} = 420\,\mbox{MeV}
\end{equation}
Unfortunately $M_2$ also contains terms which survive the
large $p$ limit. To see this one can count the number of
$\hat{p}'s$ occuring in $M_2^2$ which will give us the dominant
behaviour of $M_2^2$ for large $p$.
$\overline{V}S_0\overline{V}=M_1^4/p^2$ vanishes for large $p$
but in principle V contains a $\hat{p}$ in the nominater
$(\hat{p}A)$ and $S_0=1/\hat{p}^2$ and $\overline{V S_0 V}$
could be finite for large $p$. To be more definite consider
$$
  V S_0 V= (pA+Ap)S_0(pA+Ap)+\mbox{other terms} =
  4A_\mu{p_\mu p_\nu\over p^2}A_\nu + \ldots
$$
where we have used $[p,A_I]=i\partial_\mu A_I^\mu=0$.
$$
   \langle x|V S_0 V|y\rangle =
   4A_\mu(x)\langle x|{p_\mu p_\nu \over p^2}|y\rangle A_\nu(y) + \ldots
$$
{}From $\delta_{\mu\nu}\langle x|p_\mu p_\nu / p^2|y\rangle =
\langle x|y\rangle = \delta(x-y)\mbox{we can conclude that}
$$
  \langle x|p_\mu p_\nu/p^2|y\rangle =
  {1\over 4}\delta_{\mu\nu}\delta(x-y)$ + traceless terms.
$$
   \langle x|V S_0 V|y\rangle =
   A_\mu(x)A_\mu(x)\delta(x-y)+\ldots=\langle x|A^2|y\rangle+\ldots
$$
So $M_2^2=\overline{A^2}+\ldots=M_1^2+\ldots$ contains a term
which fully cancels $M_1^2$ calculated above. In the next section
we use an approximation which overcomes this problem but actually
restrict ourself in a calculation for small $p$.
\fi

%%%%%%%%%%%%%%%%%%%%%%%%%%%%%%%
\paragraph{5. Propagator in statistical Background (Expansion in
 Instanton Density) } \hfill
%%%%%%%%%%%%%%%%%%%%%%%%%%%%%%%

\ifall
In this section we will make an
expansion of the gluon propagator in the instanton density $n$
which will be valid for all euclidian momenta $p$ especially for
small $p$.

The average of an arbitrary power of the 1-instanton field is proportional
to the inverse volume
$ \overline{A_I^n} \sim \frac{1}{V} \quad \mbox {for} \quad n \geq 1$.
So, for example, the expansion of the square of the
multi-instanton configuration in the instanton density is
\begin{equation}
   \overline{A^2} = \sum_{I=1}^{N} \overline{A^2}
    + \underbrace{N \overline{A_I}}_{O(n)}
      \underbrace{(N-1) \overline{A_I}}_{O(n)} =
      \underbrace{N \overline{A_I^2}}_{O(n)} + O(n^2)
\end{equation}
and more general
\begin{equation}
   \overline{A^n} = N \overline{A_I^n} + O(n^2) \quad , \quad
   \overline{V^n} = N \overline{V_I^n} + O(n^2) \quad
   \mbox{for} \quad n \geq 1,
\end{equation}
where $V_I = V(A_I)$ defined in (\ref{Sinv}) with $A$ replaced
by $A_I$ and similar for $T$. Let us now sum up all terms
in (\ref{Sai}) linear in $n$:
\begin{eqnarray}\label{Sai3}
   \overline{S}^{-1} &=& (1\!\!1+\overline{T}-\overline{T^2}
    +\overline{T^3}-\ldots)S_0^{-1}+O(n^2)
\nonumber \\
        &=& (1\!\!1+N(\overline{T_I}-\overline{T_I^2}
          +\overline{T_I^3}-\ldots))S_0^{-1}+O(n^2)
\nonumber \\
        &=& (1\!\!1+N \overline{T_{eff}})S_0^{-1}+O(n^2)
         = S_0^{-1}+N \overline{V_{eff}}+O(n^2)
\end{eqnarray}
\begin{eqnarray}\label{Veff}
    T_{eff} &=& T_I-T_I^2+T_I^3-\ldots=T_I-T_I(T_I-T_I^2+\ldots)
             =T_I-T_I T_{eff}  \nonumber\\
    V_{eff} &=& V_I-V_I S_0 V_{eff}
    \quad \Longrightarrow \quad V_{eff}=S_0^{-1}(S_0-S_I)S_0^{-1}
    \quad \mbox{with} \quad \\
    S_I^{-1} &=& S_0^{-1}+V_I \quad
    \mbox{is the propagator in the 1-instanton background.}
\end{eqnarray}
More generally we can make a cluster expansion of an arbitrary
function of $A$ in the following form
\begin{equation}\label{fep}
   \overline{f(A_1+\ldots+A_N)} = \overline{f(A)} =
    \sum_{l=0}^N (^N_l)
    \overline{\sum_{k=0..l} (-)^{l-k}
    (^l_k) f(A_1+\ldots +A_k)} =
\end{equation}
$$ =\underbrace{f(0)}_{O(1)} +
    \underbrace{N\overline{f(A_1)-f(0)}}_{O(n)} +
    {1\over 2}\underbrace{N(N-1)\overline{
       f(A_1+A_2)-2f(A_2)+f(0)}}_{O(n^2)} + O(n^3)
$$
where the first line is an identity even without averaging.
Inserting a Taylor expansion for $f$ in the second line and
using the indistinguishability of different instantons one can see
that all monomials in the $k$-th term contain more or equal than $k$
different instantons. So the average will factorize in $k$ factors
each proportional to $n$ and therefore
the $k$-th term is indeed proportional
to $n^k$. It is easy to generalize (\ref{fep})
for two or more species of
fields. If $A$ is a field of $N_I$ instantons {\it and}
$N_{\bar{I}}$ antiinstantons we get to first order in $n$
\begin{equation}\label{fia}
   \overline{f(A)}=f(0)+N_I\overline{f(A_I)-f(0)}
   +N_{\bar{I}}\overline{f(A_{\bar{I}})-f(0)}+O(n^2).
\end{equation}
If we insert the propagator $S$ in (\ref{fia}) we get
\begin{equation}
  \overline{S}=\overline{S(A)}=S(0)+N\overline{S(A_I)-S(0)}+O(n^2)
   =S_0+N\overline{S_I-S_0\!}+O(n^2)
\end{equation}
which is after inversion up to $O(n^2)$ just (\ref{Sai3}).
\fi

%%%%%%%%%%%%%%%%%%%%%%%%%%%%%%%
\paragraph{6. QCD Propagators} \hfill
%%%%%%%%%%%%%%%%%%%%%%%%%%%%%%%

\ifall
In section 5 we have seen that it is enough to know the 1-instanton
propagator to calculate $\overline{S}$ to first order in $n$. Luckily
this propagator is known, unfortunatly it is a bit lengthy
and suffers from a divergency.

The gluon propagator $S_{I\mu\nu}^{ab}$ with spin $S=1$
in adjoint color\footnote{
Often called isospin \cite{Bro} }
representation $(C=1)$ can be constructed out of
the ghost propagator $\Delta_I^{ab}$ $(S=0,C=1)$
which is explicitly known in the 1-instanton background.
The general formulas how to construct a propagator of given
spin $S$ out of the corresponding scalar propagator with
same color $C$ in a selfdual background are shown below.
They are derived and more thoroughly discussed in \cite{Bro}.

{\it Notations:}
\begin{eqnarray}
   A_\mu &=& T^a A_\mu^a \quad,\quad
   [T^a,T^b] = i\epsilon_{abc} T^c \quad, \quad
   T^a = \mbox{generator of}\quad SU(2)_c
\nonumber \\
   T^a &=& \left\{ \begin{array}{rll}
           0 & \mbox{in scalar} & (C=0) \\
           \tau^a/2 & \mbox{in fundamental} & (C=\frac{1}{2}) \\
           i\epsilon_{\cdot a \cdot} & \mbox{in adjoint} & (C=1)
           \end{array} \right\} \mbox{representation}
\end{eqnarray}
$$
   P_\mu = p_\mu+A_\mu \quad,\quad p_\mu=i\partial_\mu \quad,
   \quad \tilde{G}_{\mu\nu}=\frac{1}{2}\epsilon_{\mu\nu\rho\sigma}
         G_{\rho\sigma}
   \quad, \quad \{\gamma_\mu,\gamma_\nu\}=2\delta_{\mu\nu}
$$

{\it Spin $0$ propagator $\tilde{\Delta}$:}
\begin{equation}
    \tilde{\Delta}^{-1}=P^2 \quad \Longrightarrow \quad
    \tilde{\Delta} = P^{-2}
\end{equation}

{\it Spin $\frac{1}{2}$ propagator $S$:
(not used but only stated for completeness)}
\begin{equation}
   S^{-1}= P\!\!\!\!\!\:/ = \gamma_\mu P^\mu
   \quad \Longrightarrow \quad
   S= P\!\!\!\!\!\:/ \tilde{\Delta} \frac{1+\gamma_5}{2}+
   \tilde{\Delta} P\!\!\!\!\!\:/ \frac{1-\gamma_5}{2}\quad
   \mbox{for} \quad G_{\mu\nu}= \tilde{G}_{\mu\nu}
\end{equation}

{\it Spin $1$ Propagator $S$:}
$$
   S_{\mu\nu}^{-1}= P^2 \tilde{\Delta} _{\mu\nu}+2iG_{\mu\nu}-
   (1-\frac{1}{\xi})P_\mu P_\nu \quad \Longrightarrow \quad
$$
\begin{equation}\label{glp}
   S_{\mu\nu}=q_{\mu\nu\rho\sigma}P_\rho \tilde{\Delta}^2 P_\sigma
   -(1-\xi)P_\mu \tilde{\Delta}^2 P_\nu \quad \mbox{for}
   \quad G_{\mu\nu}= \tilde{G}_{\mu\nu},
\end{equation}
$$
   q_{\mu\nu\rho\sigma} = \delta_{\mu\nu}\delta_{\rho\sigma}
   +\delta_{\mu\rho}\delta_{\nu\sigma}
   -\delta_{\mu\sigma}\delta_{\nu\rho}+\epsilon_{\mu\nu\rho\sigma}
   \quad.
$$

There are some comments in order. The above formulas are
valid for an arbitrary color representation, we will need
them only for $C=1$. For $S\neq1$
there are zeromodes and the propagator is only the inverse
of the kernel in a subspace orthogonal to the zeromodes.
The implications and problems will be discussed in section 8
when they show up explicitly. The Spin $1$ kernel is the
quadratic term of the QCD-Langrangian (\ref{LQCD}) with
gauge fixing term $\frac{1}{2\xi}(D_\mu^{ab}B_\mu^b)^2$
in slight generalisation to the $\xi=1$ case considered
previously.

With
\begin{equation}
   \Pi(x)=1+\frac{\rho^2}{x^2}\quad,\quad
   F(x,y)=1+\rho^2\frac{(\tau x)}{x^2}
    \frac{(\tau^\dagger y)}{y^2} \quad,
\end{equation}
$$
   \tau_\mu=(\vec\tau,i)        \quad,\quad
   \tau_\mu^\dagger=(\vec\tau,-i) \quad,\quad
   \tau_\mu\tau_\nu^\dagger=\delta_{\mu\nu}+i\bar\eta_{a\mu\nu}\tau_a
$$
we can write the ghost propagator for 1 instanton in the form
\cite{Bro}
\begin{eqnarray}\label{dIab}
  \Delta_I^{ab}(x,y) & = & { {1\over 2}\mbox{tr }\tau_a F(x,y)
    \tau_b F(y,x) \over 4\pi^2(x-y)^2\Pi(x)\Pi(y) } = \nonumber\\
  & = & {\delta_{ab} \over 4\pi^2(x-y)^2} -
    {\rho^2\delta_{ab} \over 4\pi^2(x^2+\rho^2)(y^2+\rho^2) } +
    {2\rho^2\epsilon_{abc}\bar\eta_{c\mu\nu}x_\mu y_\nu \over
     4\pi^2(x-y)^2(x^2+\rho^2)(y^2+\rho^2)}  \nonumber\\
  & & + {2\rho^4(((xy)^2-x^2 y^2)\delta_{ab}+
      \epsilon_{abc}\bar\eta_{c\mu\nu}x_\mu y_\nu+
      \bar\eta_{a\mu\nu}\bar\eta_{b\rho\sigma}
      x_\mu y_\nu x_\rho y_\sigma
     \over 4\pi^2(x-y)^2(x^2+\rho^2)x^2(y^2+\rho^2)y^2 }
\end{eqnarray}
For simplicity we have placed the instanton at the origin $z_I=0$
in standard orientation $O^{ab}=\delta^{ab}$.
It is possible to write down the gluon propagator
$S_{I\mu\nu}^{ab}$ explicitly in which we are finally
interested in using (\ref{glp})
but the expression will be rather lengthy.
While $\Delta_I^{ab}$ can be averaged and represented in momentum
space exactly with the help of modified Besselfunctions this seems
not to be possible for $S_{I\mu\nu}^{ab}$ and we have to expand
$\overline{S}$ for large and/or small momenta (or work much harder
to solve the complicated integrals in terms of special functions).
So we will never use the full expression for $S$.
\fi
%%%%%%%%%%%%%%%%%%%%%%%%%%%%%%%
\paragraph{7. Propagators for small momentum} \hfill
%%%%%%%%%%%%%%%%%%%%%%%%%%%%%%%

The calculation simplifies significantly if we expand
$\overline\Delta(p)$ or $\overline{S}(p)$ for small momentum.
Because $p$ always occurs in the dimensionless quantity $(p\rho)$
the lowest order in $p$ can be found by keeping only terms
of lowest order in $\rho$ or differently stated: Small $p$ corresponds
to large $x$ and for $x\gg\rho$ $\rho$ is neglegible.
To warm up let us start with $\Delta$ from (\ref{dIab}):
\begin{equation}\label{ghpr}
   \Delta_I^{ab} = \Delta_0^{ab} - \rho^2 W^{ab} + O(\rho^4)
\end{equation}
$$ \Delta_0^{ab}(x,y) = {\delta_{ab} \over 4\pi^2(x-y)^2}
$$ $$ W^{ab}(x,y) = {\delta_{ab} \over 4\pi^2 x^2 y^2} +
       {2\epsilon_{abc}\bar\eta_{c\mu\nu}x_\mu y_\nu \over
        4\pi^2(x-y)^2 x^2 y^2}
$$
$\Delta_0(p)=1/p^2$ is
the free $(A_\mu^a\equiv 0)$ ghost propagator.
After reintroducing the instanton position $z$ we can now average
$\Delta_0-\Delta_I$. The $\epsilon$-term in $W$ will be killed by
averaging over the instanton orientation:
\begin{equation}
  \langle x|\overline{\Delta_0^{ab}\!-\!\Delta_I^{ab}}|y\rangle =
  4\pi^2\delta_{ab}\rho^2{1\over V_4}
  \int {d^4\!z \over 4\pi^2(x-z)^2 4\pi^2(z-y)^2 } \quad,
\end{equation}
The last integral is infrared diverent but noticing that
$1/4\pi^2(x-y)^2$ is the fourier transformation of $1/p^2$
we can rewrite the above expression in the form
$$
  V_4 \langle x|\overline{\Delta_0\!-\!\Delta_I}|y \rangle =
  4\pi^2\rho^2\int d^4\!z \langle x|{1\over p^2}|z\rangle
  \langle z|{1\over p^2}|y\rangle =
  4\pi^2\rho^2 \langle x|{1\over p^4}| y\rangle
$$
The divergence is now hidden in the fact that the coordinate
representation of $p^{-4}$ does not exist. In fact we are not
interested in the coordinate representation but in the
momentum representation and so the divergence is spurious.
Inserting $\Delta$ in (\ref{Veff}) we get
\begin{equation}
   \overline{V}^{ghost}_{eff} =
   \Delta_0^{-1}\ \overline{\Delta_0\!-\!\Delta_I}\ \Delta_0^{-1} =
   p^2{4\pi^2\rho^2 \over V_4} {1\over p^4} p^2 =
   {1\over V_4} 4\pi^2\rho^2 = \rho^2 p^4\overline{W}
\end{equation}
and finally
$$
   M^2_{ghost}(p=0) = N\overline{V}_{eff} =
   4\pi^2\rho^2 n \approx (420\mbox{MeV})^2
$$
which has to be inserted in the ghost propagator
$\overline\Delta_{ab} = \delta_{ab}(p^2+M^2_{ghost}(p))^{-1}$.

What can we learn from this?
At first $M\not=0$. A scalar particle moving in the instanton
background gets a dynamical soft mass. The pole at $p=0$ has
vanished. Secondly $M(p\to 0)$ is finite.
reasonable to think that $M$ is bounded (has no poles) at least in the
Euclidian region $p^2>0$. From an exact averaging of (\ref{dIab})
one can see that $M$ is finit for all momenta \cite{MH}.
This was our motivation to prefer the representation
(\ref{Sai}) instead of (\ref{Sa2}). Let me make a last comment
on this inversion.
In ordinary perturbation theory one calculates
the one loop self energy, sums up the series of one loop
two consecutive loops and so on to bring the selfenergy in the
denominator of the propagator which actually is equivalent to
a simple inversion
and hopes to improve the result with this partial resummation.
The arguments for doing this are not
better or weaker than here - they are essentially the same.

Let us now calculate the gluon mass at zero momentum along the
same lines. To do this we must insert (\ref{ghpr}) into
(\ref{glp}). Expanding also $P_\mu$ up to order $\rho^2$
$$
  P_\mu=p_\mu+\rho^2\breve{A}_\mu+O(\rho^4) \quad,\quad
  \breve{A}_\mu = {2F_a\bar\eta_{a\mu\nu}x_\nu \over x^4}
$$
we get
$$
   P_\mu\Delta_I^2 P_\nu =
   p_\mu\Delta_0 p_\nu - \rho^2 p_\mu(\Delta_0 W+W \Delta_0)p_\nu +
     \rho^2(p_\mu\Delta_0\breve{A}_\nu+\breve{A}_\mu\Delta_0 p_\nu) +
     O(\rho^4)
$$
The free gluon propagator in $R_\xi$-gauge is well known or
can be obtained from (\ref{glp}) setting $G_{\mu\nu}=0$:
$$
   S^0_{\mu\nu} = q_{\mu\nu\rho\sigma}p_\rho\Delta_0^2 p_\sigma -
     (1-\xi)p_\mu\Delta_0^2 p_\nu =
   {1\over p^2}(\delta_{\mu\nu}-(1-\xi){p_\mu p_\nu\over p^2})
$$ $$
   S^0_{\mu\nu}-S^I_{\mu\nu} =
   \rho^2[q_{\mu\nu\rho\sigma}p_\rho(\Delta_0 W+W\Delta_0)p_\sigma
   -(1-\xi)p_\mu(\Delta_0 W+W\Delta_0)p_\mu]
$$ $$
   \overline{S^0_{\mu\nu}-S^I_{\mu\nu}} =
   2\rho^2\overline{W}(p^2\delta_{\mu\nu}-(1-\xi)p_\mu p_\nu) =
   2\rho^2 p^2\overline{W}S^0_{\mu\nu}
$$
where we have used in the last line that $p$ comutes with
averaged quantities like $\overline{W}$.
$$
   \overline{V}^{gluon}_{eff} = S_0^{-1}\overline{S_0-S_I}S_0^{-1} =
   2\rho^2 p^2 S_0^{-1}\overline{W}
$$ $$
   \overline{S}^{-1}=S_0^{-1}+N\overline{V^{gluon}_{eff}} =
   S_0^{-1}(1\!\!\!1+2 M^2_{ghost}(0)/p^2)
$$ $$
   \overline{S}={p^2 S_0\over p^2+M^2_{ghost}(0)} =
   {\delta_{\mu\nu}-(1-\xi){p_\mu p_\nu\over p^2} \over
     p^2+M^2_{gluon}(p^2)}\quad \mbox{with}
$$ $$
   M^2_{gluon}(p^2)=2 M^2_{ghost}(0) + O(p^2)
$$
Maybe the relation $M_{gluon}\approx 2 M_{ghost}$ is valid even for
larger $p$. Till now we have dealt with 2-color QCD $N_c=2$.
The extension of our result to the real world needs a
thorough adaption of all formulas in this work and \cite{Bro} to
$N_c=3$. After averaging there will appear several factors
$N_c/(N_c^2-1)$ like in (\ref{M1}) and $3/(N_c^2-1)$ compared
to $2/3$ and $1$ in the $N_c=2$-case. So most probable $M^2_{gluon}$
has to be multiplied with a factor between 3/8 and 1/2 yielding a gluon
mass of approximately 400MeV for $N_c=3$.

%%%%%%%%%%%%%%%%%%%%%%%%%%%%%%%
\paragraph{8. Conclusions and further developments} \hfill
%%%%%%%%%%%%%%%%%%%%%%%%%%%%%%%

In our whole calculation we have ignored the zeromodes.
All gluon field fluctuations must be orthogonal to these which
is achieved by using a gluon propagator orthogonal to the
zeromodes or otherwise stated by subtracting out from (\ref{glp})
the projections on the zeromode subspace. This procedure
also deletes a divergence comming from the second term in (\ref{dIab})
squared. There are some further problems to be solved because
the scalar product of the zeromodes with the propagator does not exist
\cite{Bro}. But all this concerns a term which is proportional
to $\rho^4$ which was irrelevant in our zero momentum approximation.
Also the principle mechanism of mass generation does not depend
on the zeromodes which can be seen from the ghost propagator
because in the spin 0 case there are no zeromodes.
This should be contrasted with the fermionic case where the zeromode
are the main ingredient for mass generation and the so called
zeromode approximation simplifies calculations enormously.

The next step may be the calculation of some gauge invariant
correlation function like a glueball correlator or topological
susceptibility. As usual this involves products and integrals
over propagators but with the difference that we have to use
(\ref{glp}) instead of the simple free propagator. Without further
simplification this may cause a headache.

\ifprivate
%%%%%%%%%%%%%%%%%%%%%%%%%%%%%%%
\paragraph{9. Private comments} \hfill
%%%%%%%%%%%%%%%%%%%%%%%%%%%%%%%

{\it
- Dim.losen Exp.par. erkl"ren

- mass cancelation from next term
- $\overline{e^S}\approx e^{\overline S}$ in section 2
  yields wrong results

- $SU(3)_c$ gluon propagator

- propagators for large $p$ $1/p^2$ coefficient
- direct calculation 1st method
- direct calculation 2st method
- comparison with OPE

- Some general integrals
- reduction to poisson equation method

- Exact spin 0 propagator for arbitrary $p$.

- Analytical continuation

- More complicated correlation functions
- testing general (ward) identities

\begin{eqnarray}
   \tilde\Delta_I(x,y) & = &
   \Pi(x)^{-1/2}{F(x,y)\over 4\pi^2(x-y)^2}\Pi(x)^{-1/2} \\
   & = & { x^2y^2+\rho^2(\tau x)(\tau^\dagger y) \over 4\pi^2(x-y)^2
     [(x^2+\rho^2)x^2]^{1/2}[(y^2+\rho^2)y^2]^{1/2} } \nonumber
\end{eqnarray}
The result is in agreement with the values calculated
by Cornwall ($500\pm 200$MeV) \cite{Cor} or
extracted from $pp$ scattering ($370$MeV) \cite{Hal}.
}
\fi
%%%%%%%%%%%%%%%%%%%%%%%%%%%%%%%
%         Bibliography        %
%%%%%%%%%%%%%%%%%%%%%%%%%%%%%%%
\ifall
\addcontentsline{toc}{section}{Literatur}
\parskip=0ex plus 1ex minus 1ex

\fi

\end{document}